\documentclass{article}

\usepackage{PRIMEarxiv}

\usepackage[utf8]{inputenc} 
\usepackage[T1]{fontenc}    
\usepackage{hyperref}       
\usepackage{url}            
\usepackage{booktabs}       
\usepackage{amsfonts}       
\usepackage{nicefrac}       
\usepackage{microtype}      
\usepackage{lipsum}
\usepackage{fancyhdr}       
\usepackage{graphicx}       
\graphicspath{{/}}   
\usepackage{ragged2e} 
\usepackage{booktabs,makecell, multirow, tabularx}

\title{Amplifying robotics capacities with a human touch: An immersive low-latency panoramic remote system
}

\author{
  Junjie Li\\
  AI-not Laboratory, School of SDIM\\
  Southern University of Science and Technology \\
  Shenzhen, Guangdong, China\\
  \texttt{lijunjie\_thu@163.com} \\
  \And
  Kang Li \\
  AI-not Laboratory, School of SDIM\\
  Southern University of Science and Technology \\
  Shenzhen, Guangdong, China\\
  \texttt{likangqd@163.com}\\
  \And
  Dewei Han\\
  AI-not Laboratory, School of SDIM\\
  Southern University of Science and Technology\\
  Shenzhen, Guangdong, China\\
  \texttt{musthan@126.com}\\
  \And
  Jian Xu\\
  AI-not Laboratory, School of SDIM\\
  Southern University of Science and Technology \\
  Shenzhen, Guangdong, China\\
  \texttt{hwllo772@gmail.com}\\
  \And
  Zhaoyuan Ma \\
  AI-not Laboratory, School of SDIM\\
  Southern University of Science and Technology \\
  Shenzhen, Guangdong, China\\
  \texttt{mazy@sustech.edu.cn} \\
}

\begin{document}
\maketitle

\begin{abstract}
AI and robotics technologies have witnessed remarkable advancements in the past decade, revolutionizing work patterns and opportunities in various domains. The application of these technologies has propelled society towards an era of symbiosis between humans and machines. To facilitate efficient communication between humans and intelligent robots, we propose the ``Avatar'' system—an immersive low-latency panoramic human-robot interaction platform. We have designed and tested a prototype of a rugged mobile platform integrated with edge computing units, panoramic video capture devices, power batteries, robot arms, and network communication equipment. Under favorable network conditions, we achieved a low-latency high-definition panoramic visual experience with a delay of 357ms. Operators can utilize VR headsets and controllers for real-time immersive control of robots and devices. The system enables remote control over vast physical distances, spanning campuses, provinces, countries, and even continents (New York to Shenzhen). Additionally, the system incorporates visual SLAM technology for map and trajectory recording, providing autonomous navigation capabilities. We believe that this intuitive system platform can enhance efficiency and situational experience in human-robot collaboration, and with further advancements in related technologies, it will become a versatile tool for efficient and symbiotic cooperation between AI and humans.
\end{abstract}


\section{Introduction}
In recent years, there has been rapid development in robotics and AI technologies, accompanied by successful commercial applications in various industries. By the end of 2022, breakthrough advancements were achieved in AIGC (Artificial Intelligence Generated Content) such as Chat-GPT and Diffusion, representing a significant leap forward in AI's capabilities in logical interactions and creativity. This has prompted a reevaluation of the collaboration models between artificial intelligence, robots, and humans, as well as the contemplation of AI replacing humans. Despite the continuous improvement in AI processing power, it still faces challenges in efficiently and effectively solving complex decision-making problems (NP-hard problems). Additionally, the demand for hardware, energy consumption, and computing power continues to rise, presenting practical difficulties in the engineering implementation of AI technologies.

To address this issue,  ``human-machine collaboration'' has emerged as a promising solution. In this approach, AI robots perform basic tasks such as grasping and navigation on behalf of humans, while humans are actively involved in complex decision-making processes, replacing AI in critical decision points. This approach effectively overcomes the challenges and high costs associated with AI's ability to handle computationally complex problems, while mitigating the risks involved in relying solely on AI for critical decision-making. To enable such collaboration, there is a need to establish a medium through which AI robots can efficiently share information with humans, allowing humans to quickly and accurately understand the on-site situation during collaborative tasks.
Remote control technology has been widely applied in various fields such as industry and security, including specialized operational robots. However, current remote-control applications commonly employ monocular or combined vision, which provides limited visual information about the operating environment. This limitation results in operators having a less intuitive and comprehensive understanding of the on-site environment and the robot's status. Additionally, mechanical delays are present when using monocular lenses in conjunction with mechanical orientation changes. These factors significantly impact the smoothness and interactive experience of remote-control operations.

Panoramic videos, capable of capturing 360° visual information, can provide operators with an immersive audiovisual experience when combined with VR devices. Moreover, the perspective shifts in panoramic videos are locally processed by VR hardware, eliminating mechanical delays and physical limitations, thus offering exceptional situational operational experiences. However, high-definition, high-frame-rate panoramic videos require high network transmission conditions\cite{lee2022real}, according to our test, 8K@60fps needs a bandwidth requirement of over 10 Mbps. Furthermore, the process of correcting and stitching videos obtained from multiple monocular cameras into panoramic videos involves encoding, streaming, network transmission, cloud forwarding (including AI intelligent processing), client streaming, decoding, and rendering. The event-to-eye latency of panoramic videos from the device to the client significantly affects the immersive remote-control experience. Previous studies have proposed optimized encoding and decoding algorithms to improve video quality and reduce latency\cite{wei2022towards,pang2019towards}. However, the achievable minimum latency has not been explicitly addressed. Research by Zheng indicates an event-to-eye latency of approximately 2.9 seconds\cite{zheng2021research}, which falls short of meeting the operational requirements in most scenarios. Tu's research indicates that they achieved an event-to-eye latency of approximately 500 ms\cite{tu2017low}.

In response to these challenges, our laboratory has proposed and developed the  ``Avatar'' platform system, aiming to establish a seamless and intuitive collaborative platform between humans and powerful AI robots. The system is built upon panoramic imaging technology, 5G network transmission, cloud-based AI algorithms, and VR hardware, creating an immersive panoramic low-latency remote operation system and platform. In this paper, we will provide a detailed introduction to the system's structure and operation in Section 2, outline the key technologies and their implementation status in Section 3, and present a conclusion and prospect in Section 4.

\section{Methods and design of the system}
In this chapter, we will provide a comprehensive overview of the overall design and concepts of the Avatar system. In addition to detailing the prototype versions of the three components (device terminal, cloud, and client) that we have implemented, we also discuss the expandability options that the system supports. These expandability options aim to accommodate a wide range of application scenarios and requirements in the era of human-machine cooperation.

\subsection{The design of the avatar system}
In a world where humans and robots coexist and collaborate, an efficient human-machine interaction platform is essential. With our developed system, humans can easily access various robots, quickly learn information about and around the robots, and supervise or assist them in making decisions. We call this system the ``Avatar System'' because when users access robots through this system, they feel as if they are immersed in the environment, and the robots become ``avatars'' for humans in on-site operations.

\begin{figure}[htbp]
	\centering
	\includegraphics[width=\textwidth]{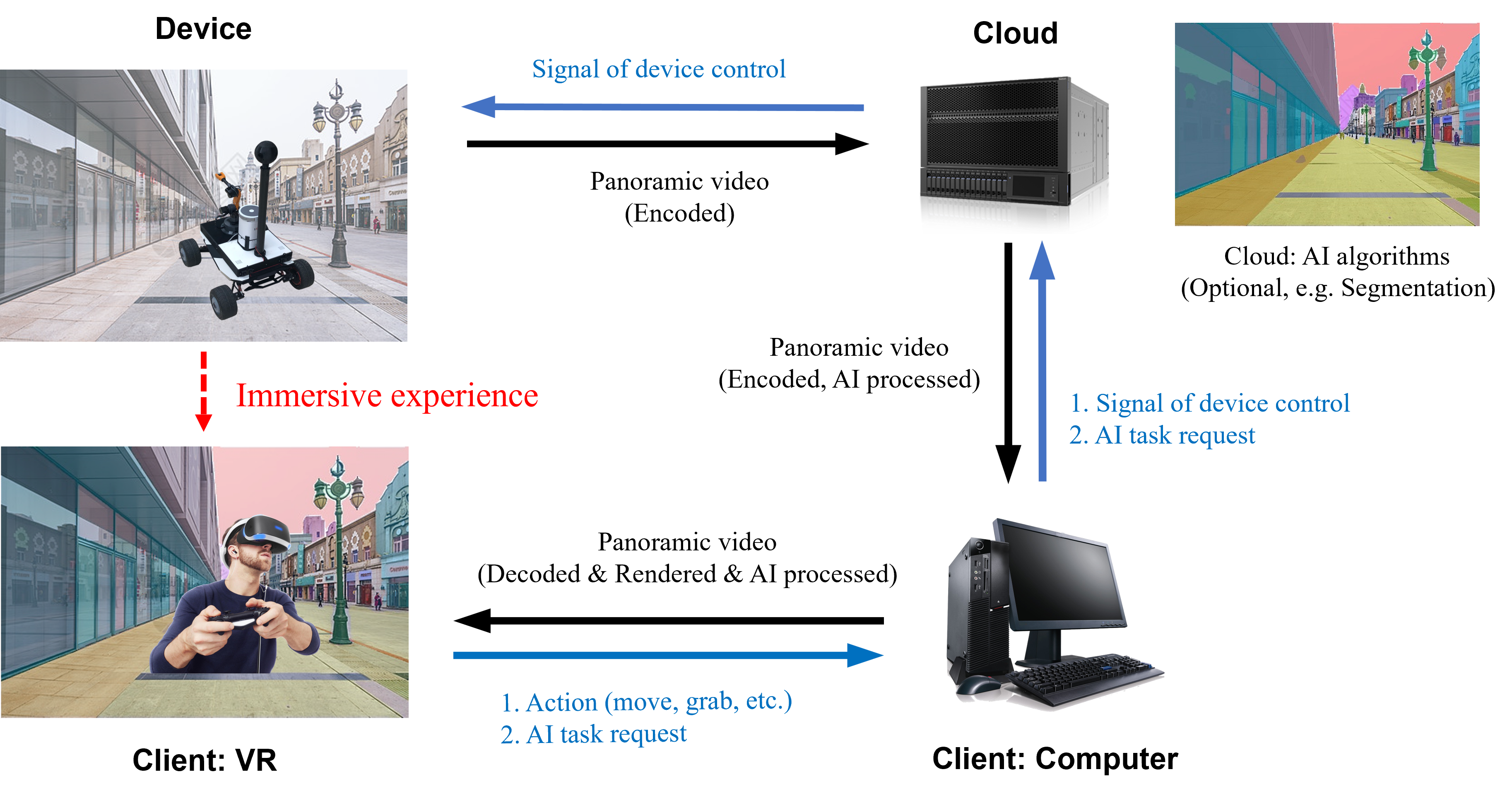}
	\caption{The components of the Avatar system}\label{1}
\end{figure}

To enable users to operate robots with minimal learning costs, we provide users with ultra-low latency (less than 200 ms) access to 360° visual and auditory information from the robots. We employ a windowed intuitive operating interface that integrates control commands for the robots and their onboard devices (such as robotic arms) into a VR visualization interface and handheld controllers. Operators do not need to know complex robot knowledge or programming commands. They can control remote robots simply by relying on intuition and simple buttons.
To achieve such functions, the Avatar System consists of device, cloud, and client components. The overall framework and operational logic are illustrated in Figure.\ref{1}. In Sections 2.2 to 2.4, we will provide a detailed introduction to the composition, functions, and implementation methods of each part.

\subsection{Device} 
The devices refer to the physical system that is remotely controlled and executes commands. Its most important functions include capturing omnidirectional video and audio, as well as fulfilling the expected functions of the terminal device. Therefore, the form of the devices can be diverse, such as various forms of robots, mobile devices, fixed location assembly line robots, and even surveillance devices. The prototype implemented in our research falls into the category of a mobile robot, but it is merely an example.

\begin{figure}[htbp]
	\centering
	\includegraphics[width=\textwidth]{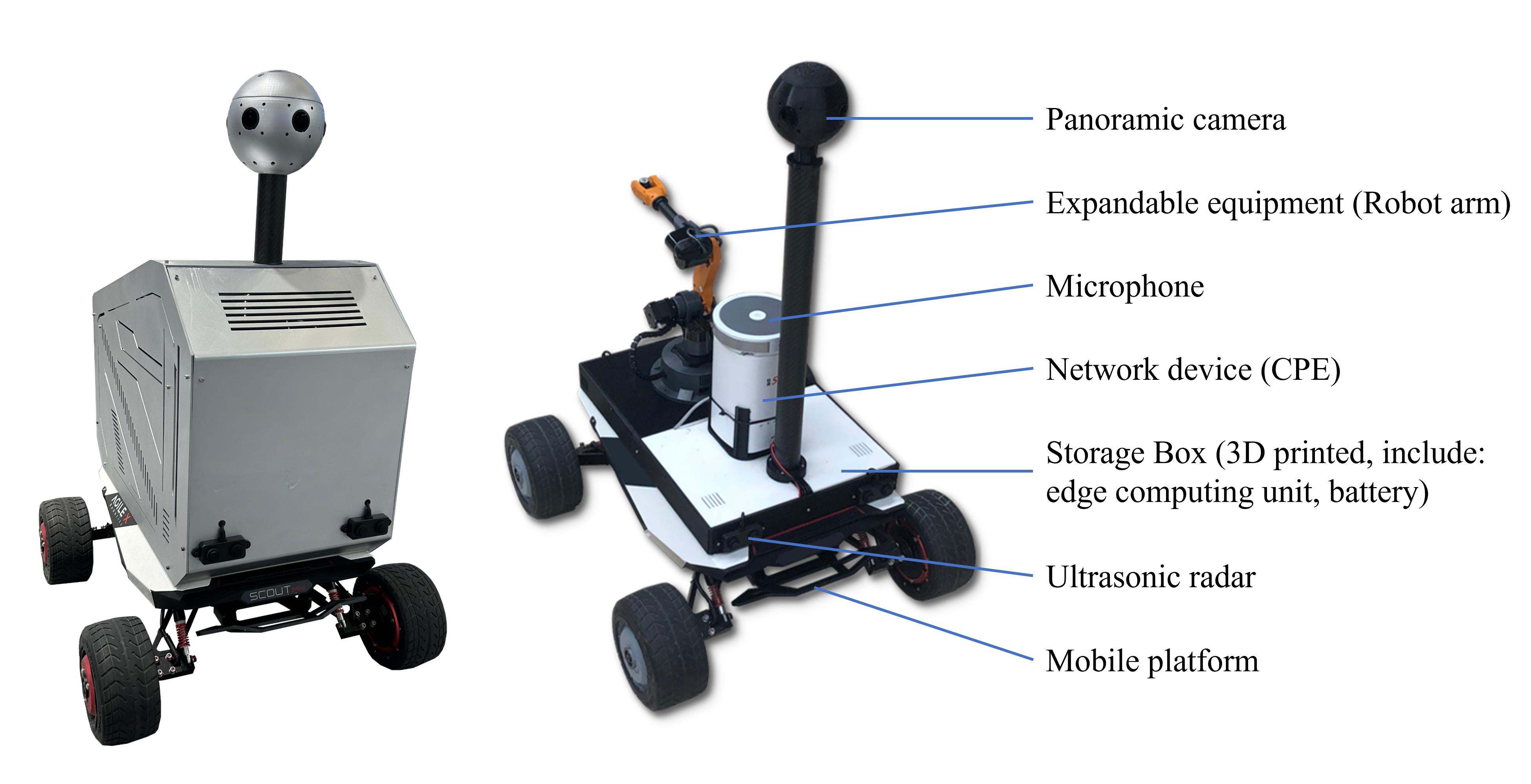}
	\caption{The prototype device of the Avatar system (without/with expandable equipment)}\label{2}
\end{figure}

In this study, our prototype takes the form of an off-road mobile platform equipped with a panoramic camera, microphone, robotic arm, edge computing unit, power battery, and network transmission device, as shown in Figure.\ref{2}. In addition to performing specific tasks with the onboard devices (such as visual grasping with the robotic arm), the main functions implemented by the devices include panoramic video capture, correction and stitching, panoramic video encoding, data transmission, and signal reception and distribution. The following section will describe the composition and functions of the main components carried by the devices.

\paragraph{Panoramic camera}
The capture of panoramic videos is achieved by a panoramic camera consisting of six fisheye lenses. Each fisheye camera has a field of view (FOV) of 120° and a focal length of 3. The six fisheye cameras are arranged in an evenly distributed manner (with an angular separation of 60° between adjacent cameras) to form the panoramic camera, ensuring a 360° coverage of the field of view as shown in Figure.\ref{2}.
\paragraph{Edge computing unit}
The prototype utilizes an edge computing unit with a computational power of only 6T. It primarily handles tasks such as distortion correction and stitching of the captured six-channel fisheye video, video encoding, and simple algorithm processing (e.g., obstacle detection using radar on the mobile platform). By employing the X method, we achieve distortion correction and automatic white balance. The stitching of the frames is performed using SIFT feature extraction and matching, resulting in a 360° panoramic RGB video. The environmental audio captured from the microphone is synchronized with the panoramic RGB video and then encoded for transmission. 
\paragraph{Network device}
We employ a CPE (model CN007) developed by China Unicom as the network transmission device. It supports the 5G network transmission protocol and provides a maximum upload and download speed of 100 Mbps. In the device terminal, it is connected to the edge computing unit via a gigabit Ethernet cable, providing network services to the edge computing unit. Its primary function is to facilitate data exchange with the cloud, including transmitting the encoded panoramic video, specific data from the onboard devices and sensors (e.g., the robotic arm in the prototype), and receiving device control signals forwarded from the cloud.
\paragraph{Mobile platform}
The prototype adopts the Agile X Scout mini as the mobile chassis. The edge computing unit receives the mobile platform control signals forwarded from the cloud and sends them to the mobile platform via the CAN protocol for execution.
\paragraph{Expandable devices}
The prototype is equipped with a six-axis robotic arm as an expandable operational device. The edge computing unit receives control signals forwarded from the cloud and sends them to the Raspberry Pi of the robotic arm for execution. By combining it with panoramic vision and cloud-based AI algorithms, the robotic arm can perform tasks such as visual grasping. The robotic arm in this study is just one form of expandable device. We aim to demonstrate through this example that the Avatar system can integrate various types of expandable devices and adapt to different terminal tasks by replacing the expandable devices and sensors based on application scenarios and requirements.

\subsection{Cloud} 
The cloud refers to the device that communicates and exchanges data with the device terminal and client through the internet or dedicated networks. It possesses functions such as data storage, data processing, and AI algorithm processing. In the Avatar system, the functionalities of the cloud server can be configured according to user requirements.

In our experiment, we used a self-built server for the cloud. It communicates with the device and the client through the internet. It has a 16-core Intel i7-11700K@3.60GHz CPU and an NVIDIA GeForce RTX 3090 GPU.

The main functions implemented in the cloud include:

\paragraph{Signaling and data transmitting}
Signaling includes two types. First, there are mobile platform control commands collected from the client, initiated by the user, and sent by the cloud to the corresponding devices. The second type is device control commands generated by cloud algorithms (e.g., intelligent navigation algorithms based on SLAM). Data includes encoded panoramic videos, sensor data, and status information from expandable devices collected from the device terminal, which are received by the cloud and forwarded to the corresponding client.
\paragraph{AI processing}
After decoding the panoramic videos, the cloud can perform various computer vision tasks based on different AI algorithms. We have implemented functions such as object detection, semantic segmentation, and human pose detection in panoramic visual images using YOLO. The processing results are annotated in the video, re-encoded, and sent to the client (the processing time of AI algorithms will contribute to the event-to-eye latency). We have also utilized ORB-SLAM3 and made improvements to utilize the front camera of the six-camera setup as the primary viewpoint, enabling map trajectory recording and autonomous navigation.
\paragraph{Data storing}
The cloud can store operational data from the devices and provide a panoramic playback feature.
In the Avatar system, the cloud is highly scalable. Users can configure GPUs, deploy AI algorithms to meet specific computing tasks, and establish archives, backups, and databases for operational data, based on their requirements.
\subsection{Client} 
\begin{figure}[htbp]
	\centering
	\includegraphics[scale=1]{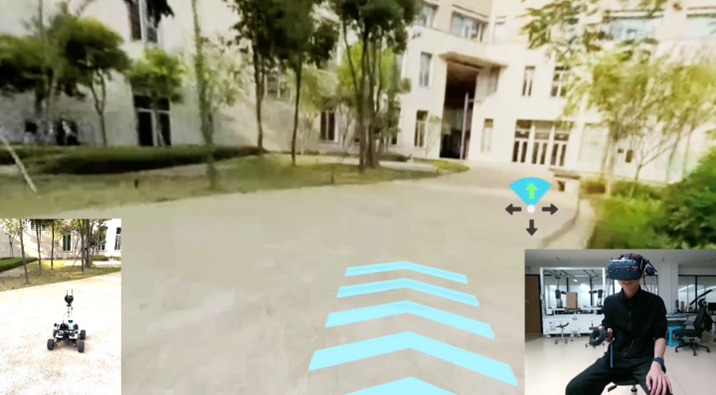}
	\caption{Operator and POV on the client of the Avatar system}\label{3}
\end{figure}

The client, which includes the hardware and software used by the operator, communicates and exchanges data with the cloud. In the proposed Avatar system, the main functions of the client include presenting the operator with real-time panoramic perception of the device and its information and collecting and sending instructions to the cloud. With the development of VR and AR hardware and technologies, the form of the client has also become diversified.
The computer used as the client can decode and render panoramic videos, providing the operator with an immersive 360° panoramic interactive experience through VR headsets and controllers. The client computer uses an i7-4790 as CPU, an NVIDIA GTX 1070 as GPU, and an HTC VIVE PRO kit as VR equipment.
Figure.\ref{3} illustrates the first-person perspective of the operator in the client and the environment where the robot is located. It can be observed that the first-person perspective in the VR headset of the operator is completely consistent with the remote robot. With the VR headset and controllers, the custom-developed client software enables real-time immersive control of the robot's movement and operation, while sharing panoramic visual and auditory information from the device terminal (remote robot) with ultra-low latency.

In the client, the main functionalities we have implemented include:
\paragraph{Immersive remote operation}
Through ultra-low event-to-eye latency panoramic videos, the client provides an immersive experience for the operator to perceive the real-time surroundings of the robot, efficiently obtain information about the robot and its environment, and remotely control the robot and its six-axis manipulator.
\paragraph{Map recording and navigation}
The prototype device is positioned as an intelligent inspection drone. With cloud-based SLAM and navigation algorithms, the client software allows recording map trajectories and enables automatic navigation.
\paragraph{Device selection}
In the Avatar system, multiple device terminals can be utilized. The client provides the capability to switch and take control of different terminals, allowing the operator to monitor the operation status of multiple devices.
\paragraph{Human-computer interaction and manual intervention}
During device intelligent operation, autonomous navigation, or other tasks, the operator can access device status, navigation maps, current position, and other information through the client interface. The operator can also intervene at any time when issues are detected or when the machine requests human assistance, helping the robot make decisions or troubleshoot problems.

\section{Experiments and Results} 
In this chapter, we will use experiments to test and validate the Avatar System's remote immersive experience capabilities. We will focus on analyzing the system's event-to-eye latency, and present the latency test results of the prototype. We also designed experiments to evaluate the operator's performance in controlling the remote robot and their response speed to environmental information when using the Avatar System.

We hope that the Avatar system can enable human-robot collaboration in the form of robots as human ``stand-ins''. In this form, the most critical link that serves as the medium for human-machine interaction is the efficient and timely information sharing between humans and machines, ensuring that humans can understand the status of robots and the surrounding environment in real-time for decision-making or assistance. Experiments by experimental psychologist Treicher showed that vision accounts for about 83\% and hearing accounts for 11\% of the ways humans acquire information from the environment. Therefore, sharing the panoramic vision and hearing of the equipment terminal can obtain about 94\% of the information on the scene, ensuring the amount of information for decision-making. Under this premise, we still need to ensure the efficiency and quality of information acquisition by operators to provide a deeply immersive feeling.

The devices used at each component of the Avatar system have been listed in Chapter 2, all of which can meet the needs of data collection, transmission, and display. In addition to the above factors, we believe that the key factor in providing deeply immersive interaction and operational experience is latency.

In the Avatar system, latency can be devided into three parts: device-to-client latency of panoramic video, latency of free perspective switching, and latency of system signal sending and execution. Among them, the data volume of system signals is small, and the encoding and decoding process is not involved in the transmission process. There is already relatively mature technical support at present, and the latency from sending instructions to device execution can be controlled below 30ms. Here we will discuss the generation, processing methods, and experimental results of the first two types of latency.

\subsection{Event-to-eye latency of panaramic vision} 
In the Avatar system using public network transmission, the processes of panoramic video from acquisition to display are as follows according to our prototype and previous research\cite{shafi2020360,yan2022dissecting}:
\begin{itemize}
  \item A single fisheye lens captures the image.
  \item The images captured by the six fisheye lenses are transmitted to the edge computing device.
  \item Image correction and stitching are performed to generate a panoramic image.
  \item The panoramic image is encoded and sent to the cloud.
  \item (Optional) Video decoding, cloud AI algorithm processing, and re-encoding of the processed results.
  \item The cloud pass data to the client.
  \item The client decodes.
  \item The client software renders and displays the panoramic video.
\end{itemize}
The fifth process is the cloud AI algorithm processing. The latency added by this process is the algorithm execution time, which will vary depending on the deployed algorithm, cloud computing performance, and other factors. The following discussion of event-to-eye latency does not include this item, that is, the cloud only forwards the encoded video from the device to the client. 

Our ultra-low event-to-eye latency is achieved by the self-designed correction and stitching algorithms of panoramic video based on the SIFT feature, H.265(HEVC) format for encoding and decoding, and 5G network transmission. 

In the Avatar system, we define the event-to-eye latency of panoramic video as the time difference between the moment the event is captured at the device and the moment the user sees the event in the VR glasses at the client. Based on this definition, we tested the latency of the Avatar system under good network conditions.

\begin{figure}[htbp]
	\centering
	\includegraphics[width=\textwidth]{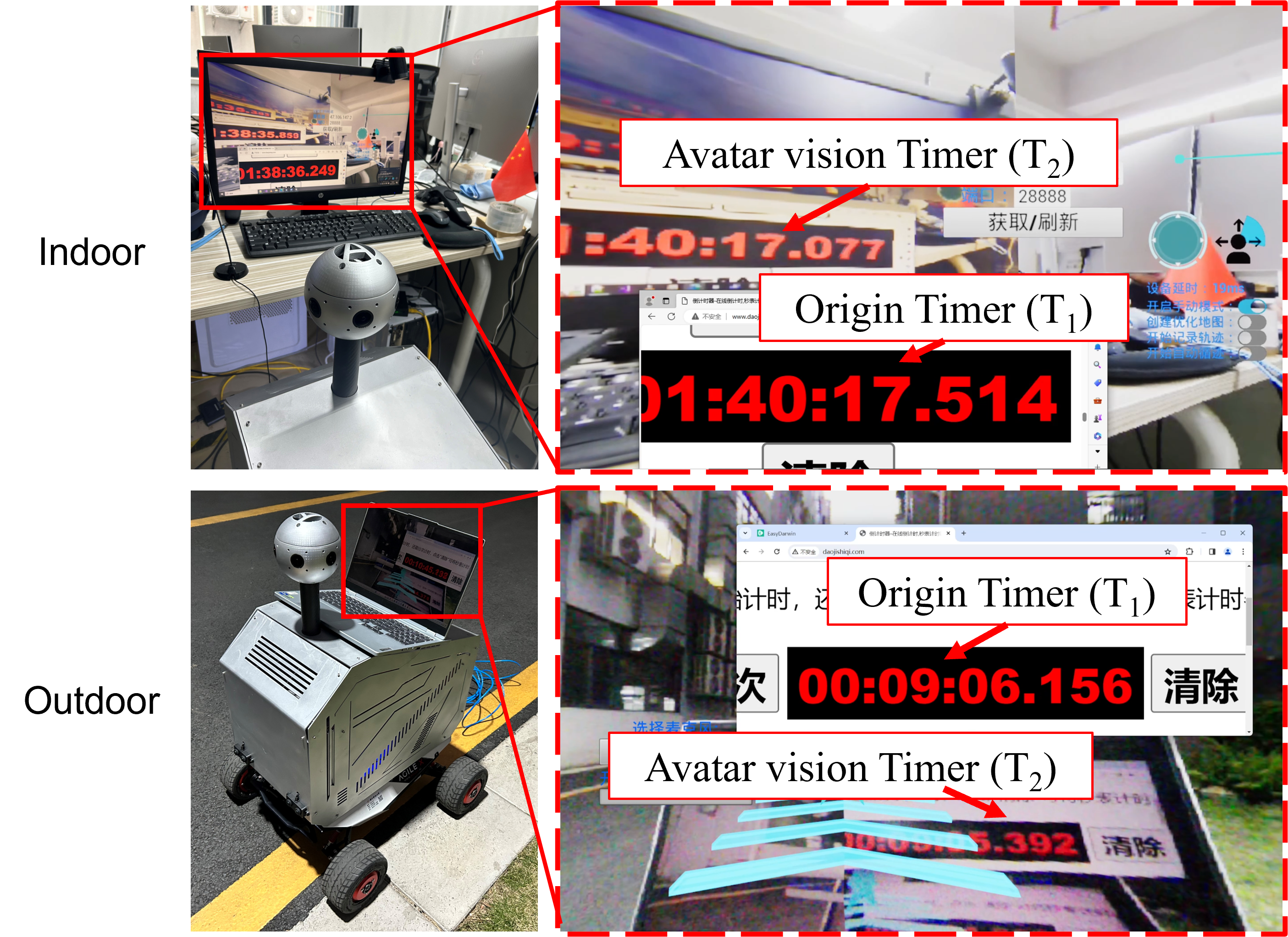}
	\caption{Test of event-to-eye latency in the Avatar system}\label{4}
\end{figure}

The experimental method for testing the latency of the prototype: use the panoramic camera at the device to shoot the time T1 on the computer screen of the client, and adjust the perspective in the VR glasses until you can see the screen time T1. Synchronize the screen of the VR glasses to the computer screen for display, and get the time T2 displayed on the client. At this time, the screen simultaneously has time T1 and T2, T1 is the origin timer, and T2 is the Avatar vision timer after processing by the Avatar system device, cloud, and client. The time difference between T1 and T2 is the event-to-eye latency of panoramic video. Since T1 and T2 are constantly changing during system operation, for easy calculation, we take a screenshot of the client screen record to obtain the value at a certain moment. The test equipment and process are shown in Figure.\ref{4}.

We performed five event-to-eye latency tests on the Avatar system at three different locations (laboratory, office and outdoor) using the above method. Each test lasted 60 seconds, starting from the 5th second of the 60-second video, and taking a screenshot every 5 seconds to record T1 and T2. Each test recorded 12 differences T= T1- T2, and the average value was calculated after removing the highest and lowest values. Our recording results are shown in Table ~\ref{tab:table1}.


\begin{table}[h]
 \caption{Results of event-to-eye latency test}
  \centering
  \begin{tabular}{lccc}
    \toprule
    Location & Test No. & Average latency (s)  & Standard deviation (s) \\
    \cmidrule(r){1-4}
    \multirow{6}*{Laboratory} 
    & 1 & 0.4649 & 0.033398 \\ 
    & 2 & 0.4418 & 0.036599 \\ 
    & 3 & 0.4400 & 0.012156 \\ 
    & 4 & 0.4418 & 0.035219 \\ 
    & 5 & 0.4597 & 0.040244 \\ 
    & Average & 0.44964 & 0.031523 \\
    \cmidrule(r){1-4}
    \multirow{6}*{Office} 
    & 1 & 0.3537 & 0.029560 \\ 
    & 2 & 0.3688 & 0.022929 \\ 
    & 3 & 0.3559 & 0.031859 \\ 
    & 4 & 0.3463 & 0.006129 \\ 
    & 5 & 0.3616 & 0.022505 \\ 
    & Average & 0.35726 & 0.022596 \\
    \cmidrule(r){1-4}
    \multirow{6}*{Outdoor} 
    & 1 & 0.8138 & 0.062641 \\ 
    & 2 & 0.7675 & 0.044702 \\ 
    & 3 & 0.8070 & 0.067431 \\ 
    & 4 & 0.7848 & 0.032832 \\ 
    & 5 & 0.8071 & 0.051416 \\ 
    & Average & 0.79604 & 0.051805 \\
    \bottomrule
  \end{tabular}\label{tab:table1}
\end{table}

As can be seen from Table~\ref{tab:table1}, the average event-to-eye latency of the Avatar system is 449.64 ms in the laboratory, 357.26 ms in the office, and 796.04 ms outdoors. The office environment has the best network conditions, it has the lowest event-to-eye latency and the smallest standard deviation. The outdoor environment has the worst network conditions, so the latency is the highest. The standard deviation of the latency is relatively small indoors, indicating that the latency is relatively stable. But it becomes a little higher outdoors, which means there exist some fluctuations. 

In general, the event-to-eye latency of the Avatar system is within a reasonable range, especially in stable network environments like laboratory and office, which can meet the requirements of human-machine collaboration in most scenarios. 

\subsection{Latency of perspective switching}
The latency of perspective switching is part of event-to-eye latency. In non-panoramic vision schemes, when the perspective is switched, the mechanical structure is needed to change the direction of the camera. This scheme will cause the latency of perspective transformation due to the delay of mechanical movement, which will hurt the immersive experience.

In our system, the panoramic camera at the device end continuously captures 360° panoramic images in real-time. The client obtains and decodes them, and then renders them in real time for panoramic video display at the VR end. The VR hardware adopts a head-tracking-based perspective switching method, which does not need to exchange additional data with external devices, but only needs to update the perspective according to the change of user head movement\cite{yaqoob2020survey}. The head-mounted display (HMD) worn by the user is equipped with sensors to track the user's head movement (including the translation and rotation of the user's head), which is used by the VR system to calculate the new perspective. The new perspective data is transmitted to the HMD for display screen update. Since the VR system internally transmits this type of data very quickly, the calculation of the field of view direction based on the head posture can be completed in real-time, completely free from the mechanical delay or physical limitations of the remote robot. Therefore, the prototype we implemented is completely free and almost zero latency in perspective switching.

\section{Conclusion}
In this paper, we propose a general-purpose platform for the era of human-machine symbiosis, called the Avatar System. The core concept of the Avatar System is to allow humans to efficiently share information in a way that fully complies with human intuition and perception habits by sharing the perception of a robot with low latency and panoramic audiovisual sharing. To this end, we have built an Avatar System prototype in the form of devices, cloud, and clients, and introduced the functions and implementation methods of each end. In the experimental part, we tested the event-to-eye latency that directly affects the perception effect of humans and explained the implementation and advantages of the system's ultra-low latency in free-viewpoint transformation.

As can be seen from the experimental results, the prototype we built can complete multiple basic functions such as low-latency perception, remote operation, and human-machine cooperation, which is an engineering realization of the concept of the ``Avatar'' system. The conditions for the implementation of the prototype system include panoramic visual acquisition devices, a transmission network with sufficient bandwidth, and a VR client that can realize immersive panoramic and remote operations. In the follow-up research, we will further optimize the event-to-eye latency, and focus on improving the system's interaction experience. In the future, we will design reasonable evaluation methods and experiments to quantify the experience of the system in terms of immersion and interaction, and continuously improve the system based on this.

The system is expected to be applied in scenarios such as dangerous operations, remote interaction, and health care and nursing. For example, by connecting the Avatar System to mining equipment, chemical machinery, inspection robots, and nursing robots, workers and doctors can improve their work efficiency and ensure their safety. In recent years, AI algorithms have made significant breakthroughs, and metaverse technology has continued to develop. It can be foreseen that these technologies will continue to provide upgrades for the system, such as expanding device-side functions and optimizing client-side interaction forms. This is expected to further improve the efficiency and experience of human-machine cooperation, and further expand the application fields of the Avatar System.

We believe that the Avatar System can play an effective role as a medium in the application field of artificial intelligence and robotics technology. With the strategy of human assistance and monitoring, it can provide new ideas and low-cost implementation methods for the application of non-general AI in industry, life, and other fields, and accelerate the process of AI and robotics technology benefiting mankind.

\section*{Acknowledgments}
The authors are appriciated for the supports by Southern University of Science and Technology for testing venue and hardware, and the CPE equipment support by China Unicom. We are also grateful for the algorithms support by the open source community on the internet like ORB\-SLAM3\cite{campos2021orb}. 

\bibliographystyle{unsrt}  
\bibliography{references}

\end{document}